# Low-dose Chemically Specific Bioimaging via Deep-UV Lensless Holographic Microscopy on a Standard Camera

Piotr Arcab[1,*], Mikołaj Rogalski[1], Karolina Niedziela[1], Anna Chwastowicz[2,3], Emilia Wdowiak[1], Julia Dudek[1], Julianna Winnik[1], Paweł Matryba[2,3], Jolanta Mierzejewska[4], Małgorzata Lenarcik[5,6], Ewa Stępień[7], Piotr Zdańkowski[1], Grzegorz Szewczyk[8], and Maciej Trusiak[1,**]

[1]Institute of Micromechanics and Photonics, Faculty of Mechatronics, Warsaw University of Technology, Warsaw, Poland
[2]Department of Immunology, Medical University of Warsaw, Warsaw, Poland
[3]Laboratory of Neurobiology, Nencki Institute of Experimental Biology of Polish Academy of Sciences, Warsaw, Poland
[4]Warsaw University of Technology, Chair of Drug and Cosmetics Biotechnology, 3 Noakowskiego St., 00-664 Warsaw, Poland
[5]Department of Pathology, Maria Sklodowska-Curie National Research Institute of Oncology, Warsaw, Poland
[6]Department of Gastroenterology, Hepatology and Clinical Oncology, Centre of Postgraduate Medical Education, Warsaw, Poland
[7]Department of Medical Physics, M. Smoluchowski Institute of Physics, Faculty of Physics, Astronomy and Applied Computer Science, Jagiellonian University, Krakow, Poland;
[8]Faculty of Biochemistry, Biophysics and Biotechnology, Jagiellonian University, Krakow, Poland

*piotr.arcab.dokt@pw.edu.pl
**maciej.trusiak@pw.edu.pl

**Abstract**

Deep-ultraviolet (DUV) microscopy can provide label-free biochemical contrast by exploiting the intrinsic absorption of nucleic acids, proteins and lipids, offering chemically specific morphological information that complements structural optical thickness contrast from phase-sensitive imaging. However, existing DUV microscopes typically rely on specialized optics and DUV-sensitive cameras, which restrict field of view, increase system complexity and cost, and often require high illumination doses that risk photodamage. Here, we report a low-dose deep-UV lensless holographic microscopy platform that uses standard board-level CMOS sensors designed for visible light, eliminating all imaging optics and dedicated DUV detectors. Our system achieves large field-of-view (up to 116 mm²) DUV imaging with low illumination and label-free phase and chemically specific amplitude contrast. A specialized defocus/wavelength diverse pixel super-resolution reconstruction with total-variation regularization and robust autofocusing halves the effective sensor pixel pitch and yields down to 870 nm lateral resolution. We demonstrate chemically specific, label-free bioimaging on challenging specimens, including *Saccharomyces cerevisiae*, extracellular vesicles and unstained mouse liver tissue. In liver sections, imaging at $\lambda = 330$ nm reveals lipid- and retinoid-rich accumulations that co-localize with Oil Red O staining, enabling label-free identification of hepatic stellate (Ito) cells. This combination of low-dose operation, chemically specific contrast and standard CMOS hardware establishes DUV lensless holographic microscopy as a practical and scalable route to high-content submicron-resolution whole-slide preparation-free bioimaging without exogenous labels.

## Introduction

High-resolution, high-throughput, label-free imaging of transparent biological specimens, e.g., cell cultures and tissue slices, remains one of the central challenges in optical microscopy[1,2]. Conventional bright-field microscopes operating in the visible (VIS) spectral range often fail to produce sufficient image contrast when dealing with unstained cells and tissues. Most cellular structures are of low absorption, which results in minimal intensity modulation and poor visibility of crucial fine features[1,2]. As a consequence, a wide range of biological studies still rely on the use of extrinsic labels, such as absorbing dyes or fluorescent markers, which enhance contrast and allow for identification of specific structures. However, these labelling protocols are time-consuming, prone to variability, and often perturb the native state of the specimen[1,2]. In addition, labelling introduces chemical and procedural complexity, increasing cost and limiting throughput in biomedical workflows.

A major breakthrough in label-free imaging has been achieved through Quantitative Phase Imaging (QPI) techniques[1,2]. QPI provides high-contrast visualization of transparent samples by mapping variations in refractive index and physical thickness. In contrast to intensity-based methods, QPI can reveal fine structural details without the need for external markers, which enables long-term observation of living cells in their physiological environment without photodamage. Many implementations have been developed, including high-throughput digital holographic microscopy[1,2,3] or high-resolution Fourier ptychography[4]. These techniques have opened powerful opportunities in cell biology and tissue pathology imaging. However, QPI has a fundamental limitation - it lacks intrinsic chemical specificity. Since it relies solely on optical thickness (refractive index and physical thickness) contrast, QPI provides structural information but cannot distinguish between different biochemical components. In many biomedical and diagnostic applications, this lack of molecular sensitivity restricts its usefulness.

To address this limitation, researchers have increasingly focused on Deep Ultraviolet (DUV) imaging[5-10]. DUV offers unique advantages for label-free biochemical contrast[11-13]. Biomolecules such as nucleic acids and proteins exhibit strong intrinsic absorption bands in the DUV spectral range, e.g., DNA absorbs strongly at wavelength of approximately 260 nm, while lipids absorb around 320 nm[14,15]. This native absorption can be used to provide label-free chemically specific contrast that effectively acts as a built-in molecular stain. In addition, the short wavelength of DUV light can lead to improved diffraction-limited resolution compared to the visible range, which enables submicron structural visualization. These two factors, chemical specificity and improved resolution, make DUV a very attractive modality for biomedical imaging and histopathology.

Practical implementation of DUV microscopy is technically challenging[16], however. Standard glass optics strongly absorb DUV radiation. Only specialized materials such as fused silica or reflective components, such as Cassegrain objectives, can be used to construct optical imaging systems. High-quality DUV objectives are expensive. Even with advanced materials, their numerical aperture (NA) remains relatively low, which limits the achievable resolution. Furthermore, traditional lens-based architectures inherently restrict the field of view (FOV). Imaging extended tissue sections or large cell cultures efficiently is, therefore, cumbersome. These constraints, including high cost, limited NA, small FOV, and potential chromatic aberrations, have so far limited the widespread use of DUV in practical biomedical imaging. DUV illumination at high intensities can also induce phototoxic effects, which further complicates live-cell studies. Recently, Gorti et al.[17] quantified UV-induced photodamage and reported a critical total fluence threshold for significant cell death of approximately 1 µJ/µm². Under reported DUV imaging conditions (plasma-laser source at a wavelength of 255 nm, power density ~1 nW/µm², 50 ms exposure time), the fluence per frame is around 0.05 nJ/µm², allowing the authors to perform more than 20,000 sequential acquisitions before reaching the damage threshold.

A promising strategy to bypass these constraints is to adopt a lensless imaging approach. Lensless digital holographic microscopy[7,12,18,19] (LH) removes lenses entirely and directly records the diffracted optical field as a hologram on a sensor. The complex sample information is then numerically reconstructed by computational back-propagation to the sample plane. By eliminating conventional optics, LH avoids the absorption and aberration issues that are present in DUV lens-based systems and is thus a perfect candidate for DUV imaging development. This trend is echoed by an alternative lensless ptychographic

UV imaging approach, which further illustrates the growing interest in DUV-enabled label-free modalities[20]. Lensless holography significantly reduces cost and system complexity[21–23]. In addition, standard CMOS cameras originally designed for the visible range provide sufficient sensitivity in the DUV band. This eliminates the need for costly UV-optimized detectors and further reduces the overall system cost without compromising imaging performance. Because the sample is placed in close proximity to the sensor (around 1-2 mm), the effective FOV is determined only by the sensor area. Such a large FOV is particularly beneficial for high-throughput applications such as digital pathology, cytology, and large-area cell/tissue screening[24,25]. Importantly, LH has also demonstrated excellent performance under low photon illumination conditions, maintaining high phase reconstruction quality[26,27]. This highlights the intrinsic efficiency of lensless architectures, which require significantly lower illumination intensities compared to conventional optical setups, thereby minimizing phototoxicity.

In this work, we present a straightforward and cost-effective platform for Lensless Holographic Deep Ultraviolet (LH-DUV) microscopy. This label-free system integrates a new robust computational pixel super-resolution[28] (PSR) with total-variation (TV) denoising[29] and darkfield autofocusing to achieve large FOV, high-contrast, molecularly sensitive imaging of biological samples under low illumination and deploying regular camera. Because DUV absorption is intrinsically tied to specific molecular bonds, LH-DUV simultaneously provides amplitude contrast with inherent biochemical specificity and complementary quantitative phase imaging. To demonstrate the capabilities of LH-DUV bioimaging, we validated the system on yeast cells, extracellular vesicles (EVs), and mouse liver tissue. This novel platform combines wide-area imaging, low photon budget operation, chemical specificity, and low cost (few optomechanical parts and a regular VIS camera), which makes it a promising tool for applications in digital pathology, cytology, and life sciences.

**Results**

Experimental Setup and Reconstruction Strategy

The experimental platform is based on a simple LH configuration designed to enable cost-effective, large FOV imaging in the DUV regime (Fig. 1). The sample was positioned as close to the sensor as it could be (approximately 1-2 mm, depending on measured sample) to achieve ~1x magnification (with FOV equal to the camera size), and plane-wave illumination was used. The scattered and unscattered components of a wavefront interfere directly on the detector, producing an in-line Gabor hologram. In the simplest configuration, this hologram can be numerically back-propagated to the sample plane, e.g., using the angular spectrum[30] (AS) method. However, such reconstruction would be spoiled by reconstruction artefacts[31] due to missing phase information in camera plane and lateral resolution limited by the pixel size of the used camera.

In our study, we employed a data multiplexing strategy that enables to simultaneously minimize the twin image artefacts and bypass the pixel size resolution limitation with the PSR algorithm[28]. The employed experimental system was capable to work with two equivalent data multiplexing modes: (i) axial multi-distance (various sample-camera distances) and (ii) multi-wavelength (different illumination wavelengths). Both approaches provide the hologram-domain sub-pixel diversity required to surpass the native sensor pitch that limits sampling-driven resolution; data acquired with both diversity methods were processed using the same PSR reconstruction procedure. Importantly, the employed reconstruction algorithm was adapted to low intensity (low signal-to-noise ratio, SNR) holograms - which can result due to DUV source and VIS camera spectral mismatch and minimizing exposure time to reduce photodamage - by additionally supporting it with TV. The proposed reconstruction procedure is described in details in the Methods section. A representative hologram (blue frame in Fig. 1(a)) and its reconstruction (Fig. 1(b)) resolving fine mouse liver tissue features are shown (in amplitude) in Fig. 1.

Illumination was provided by a tunable nanosecond pulsed laser, enabling direct comparison between VIS and DUV reconstructions. For one validation experiment (shown in Fig. 2), a fixed-wavelength

DUV laser was also employed to demonstrate that the method can operate effectively with simpler and lower-cost light sources when employing an axial multi-distance diversity configuration. This highlights that while spectral tunability facilitates systematic wavelength-dependent analysis, the core lensless configuration remains compatible with standard single-wavelength DUV illumination.

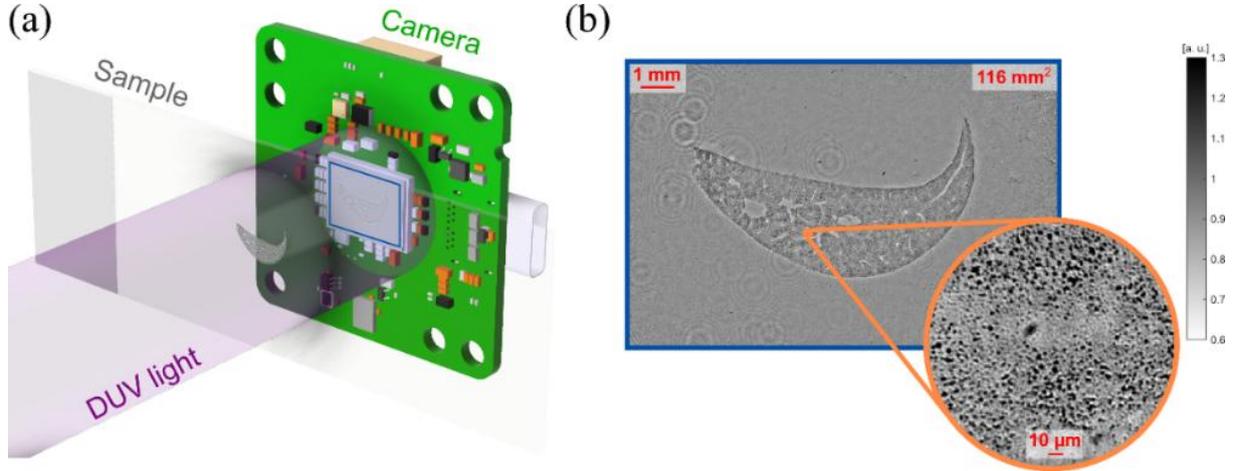

Figure 1. (a) Scheme of the LH-DUV setup. The sample is illuminated by a plane wave, and the camera is placed in close proximity to the sample (~1–2 mm). Unscattered and scattered components of the transmitted field interfere directly on the sensor, forming an in-line Gabor hologram. Multi-frame acquisition for PSR reconstruction is achieved either by varying the sample–camera distance through axial camera displacement or by tuning the illumination wavelength of the laser source. (b) Representative PSR amplitude reconstruction of a mouse liver section hologram recorded using 340 nm wavelength, illustrating the large accessible FOV and resolving fine DUV-specific histological features.

## Low-photon budget operation in DUV with VIS-specified cameras

To validate that CMOS cameras originally designed for the visible range can also operate in the LH-DUV, and to highlight the inherent capability of LH-DUV to maintain resolution even under extremely low illumination[26,27], we conducted a series of experiments with controlled exposure times using a board-level VIS camera with 1.45 μm pixel size. In the DUV case, the optical power reaching the detector (camera plane) was below the sensitivity of any of our available power meters (<100 μW), underlining the low-light conditions of the experiment. Because no direct measurement of such weak power levels was possible, we required an alternative way to control and quantify the effective illumination. To this end, we kept the illumination constant and systematically reduced the camera exposure time. In this manner, we were able to indirectly emulate progressively lower doses of incident light, while ensuring that the level of signal recorded by the sensor was well controlled. This approach allowed us to demonstrate that reconstructions with maintained spatial resolution can still be achieved from data acquired at the very limit of detectable hologram intensity. DUV measurements in this section were performed at $\lambda = 266$ nm, and the data used for PSR were acquired in the multi-distance mode.

Figure 2 presents four datasets (rows) corresponding to different camera exposure times (0.3, 0.5, 1, and 15 ms), each reconstructed using three distinct algorithms. The first column shows Gabor reconstructions obtained via the AS method, which (contrary to PSR) was performed from only a single hologram. The second column presents PSR reconstruction results, and the third column displays PSR results further enhanced with TV and autofocusing. The last column illustrates the effective bit depth of a single hologram for each representative exposure dataset. As shown, the application of PSR consistently improves resolution, while additional TV denoising effectively suppresses noise. A detailed quantitative analysis of the denoising influence is provided in the Supplementary Material, wherease proposed PSR with TV and autofocusing algorithm is carefully described in the Methods section.

Even when the recorded hologram contained only three gray levels (first row), the final reconstruction of amplitude USAF test target (using PSR with TV) achieved a spatial resolution of 1.74 μm (Group 8, Element 2). A slight increase in signal - corresponding to four gray levels (second row) - enabled a resolution of 0.98 μm (Group 9, Element 1). Further increases in exposure time (third row), led to a final resolution of 0.87 μm (Group 9, Element 2), matching the resolution obtained under long-exposure conditions, where holograms contained approximately 150 gray levels (fourth row). This indicates that the proposed PSR with TV reconstruction pipeline is capable of preserving high spatial resolution and delivering robust SNR even for extremely low-photon-budget DUV holograms - an important advantage given both the camera–source spectral mismatch in the DUV and the need to minimize illumination levels to bypass phototoxicity in biological samples.

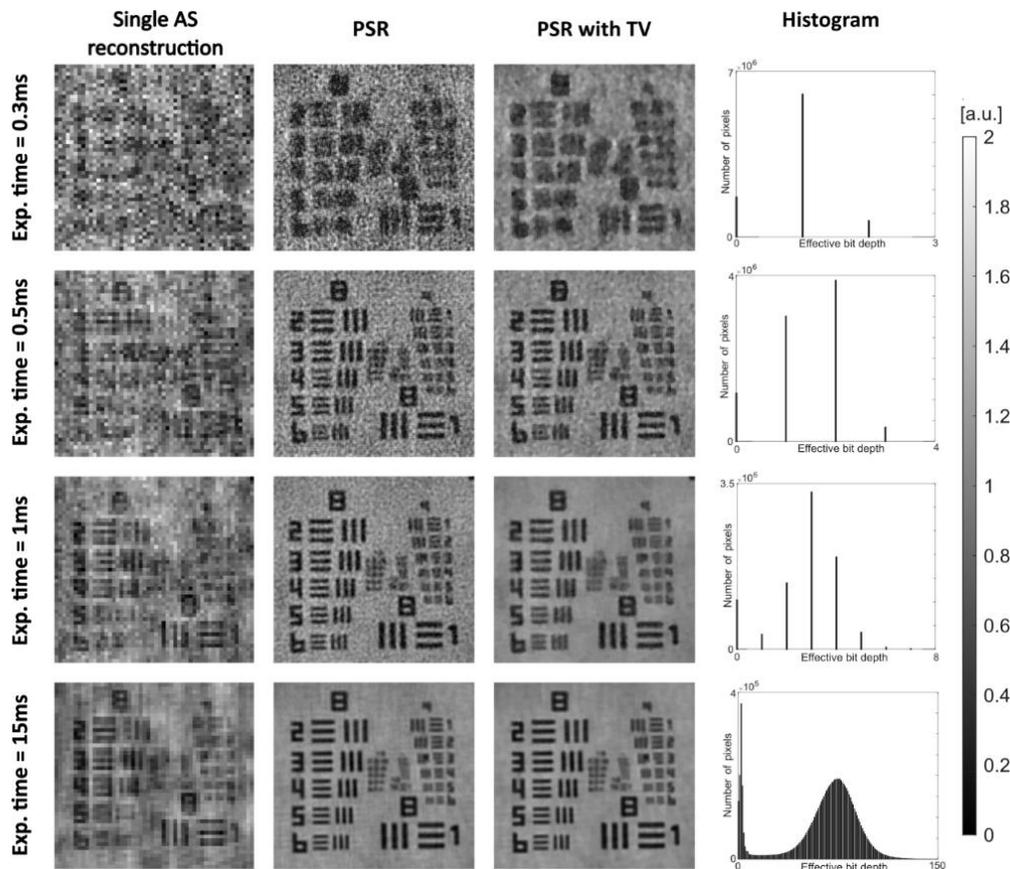

Figure 2. LH-DUV reconstructions of the finest resolved features of a USAF resolution target. Four datasets (rows) were acquired with different exposure times to modulate the recorded signal level. The first row corresponds to the minimal exposure enabling reconstruction, yielding only three effective gray levels. Slightly increasing the exposure (second row) produces four gray levels and a clear improvement in the achievable resolution. Further increasing the illumination (third row) results in several gray levels and achieves the same resolution as obtained under standard recording conditions (fourth row). Each dataset was reconstructed using three methods: single-frame AS back-propagation (first column), multi-frame PSR (second column), and the proposed PSR enhanced with a TV denoising step optimized for extremely noisy data (third column). The last column reports the effective bit depth of each dataset.

Phase-amplitude contrast in VIS vs. DUV lensless holographic microscopy

DUV illumination is expected to enhance image contrast for most samples, which are otherwise transparent in visible range, due to significantly higher absorption at DUV wavelengths. Additionally, the accumulated phase shift scales as $\varphi \sim \frac{1}{\lambda}$, so decreasing λ enhances overall phase sensitivity. For submicrometer features, the scattering efficiency increases toward the DUV and reaches a maximum when the wavelength approaches the feature size, increasing the in-line holographic signal. Importantly, many biomolecules and materials exhibit strong and spectrally distinct absorption bands in the DUV. This wavelength-dependent absorption provides a powerful source of intrinsic chemical contrast, effectively acting as a label-free molecular "stain." Together, these effects can reveal structures that are weak or invisible in the VIS range and can convert purely phase contrast into mixed phase–amplitude contrast. To quantify these effects and assess how PSR facilitates the recovery of fine phase details, we used a custom phase resolution target (Lyncée Tec, Boroflat 33 glass, height 125±5 nm).

The top panel of Fig. 3 shows the full FOV phase reconstruction of the phase resolution target. The pink square indicates the region that was analyzed in greater detail, shown on the right. A single-frame reconstruction obtained using the AS method resolves the smallest designed features of the target, whereas the corresponding PSR reconstruction additionally reveals submicron structures, such as small debries particles (marked with arrows), that remain invisible in the AS reconstruction alone. In this experiment, PSR was implemented in the multi-wavelength mode. For both VIS and DUV measurements, 22 holograms were recorded with a wavelength step of Δλ = 3 nm. The VIS sweep covered λ = 547–610 nm, while the DUV sweep spanned λ = 238–301 nm.

As expected for a purely phase object, the VIS measurements (shown below the FOV) exhibit strong contrast in the phase channel, while the amplitude shows only a contrast weak modulation. Contrary, DUV illumination renders the same structures clearly visible in both phase and amplitude, reflecting the increased amplitude and scattering, and higher phase sensitivity at shorter wavelengths. This confirms that transparent features effectively phase-only in the VIS acquire measurable amplitude contrast in the DUV. The structures adjacent to the letter "S" correspond to a 2 μm linewidth, which is the smallest element on our phase target. The ultimate resolution of the system is analyzed in the following section.

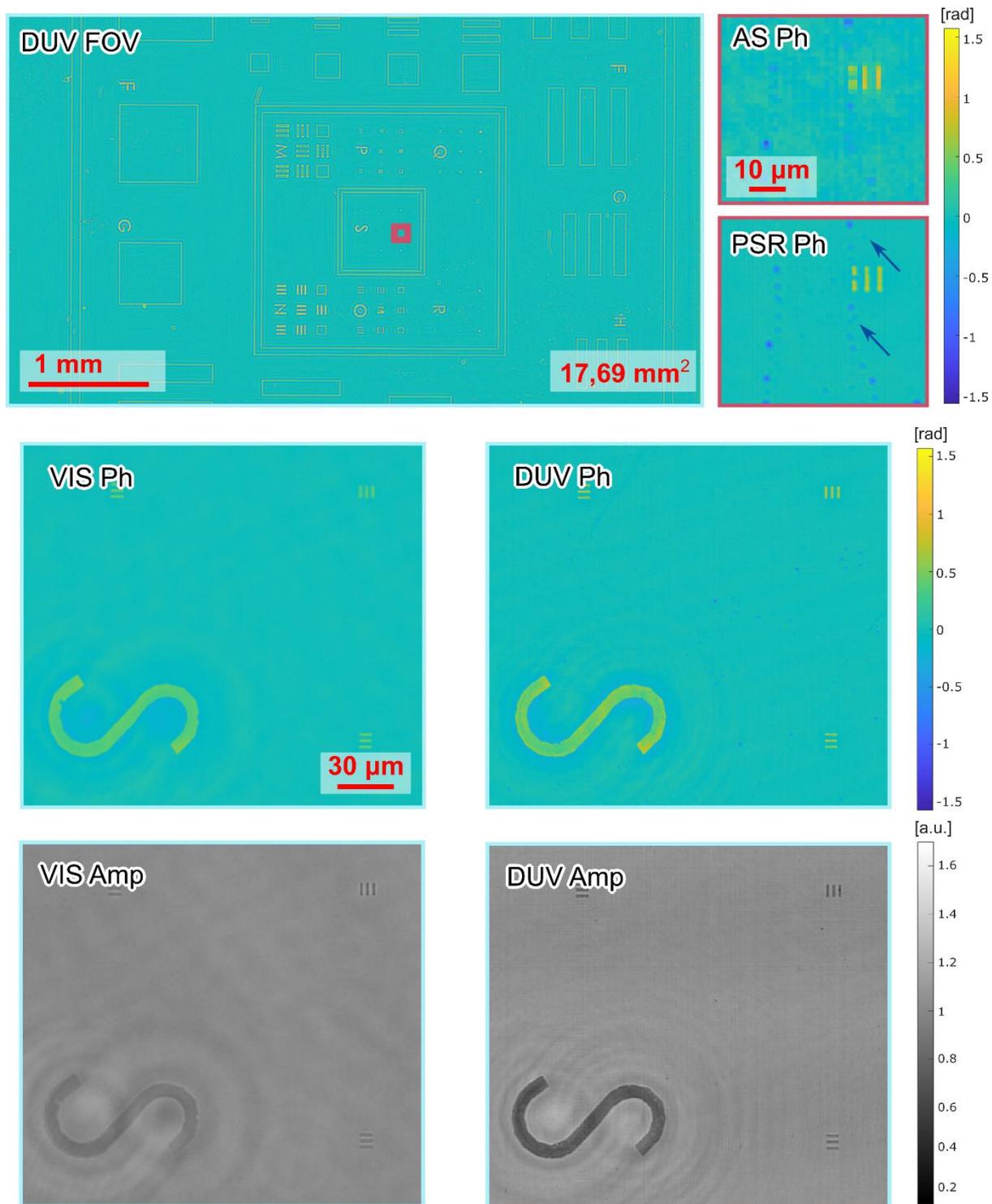

Figure 3. The top panel shows the full FOV of a phase test target recorded with the LH-DUV system. The pink square marks the region enlarged on the left, where the same area is reconstructed (in amplitude Amp and phase Ph) using single-frame AS back-propagation (upper ROI) and multi-frame PSR (lower ROI). The PSR reconstruction reveals phase features that are not resolved in the AS result. The panel below FOV presents a comparison of VIS and DUV reconstructions in amplitude and phase. In the DUV regime, phase variations partially leak into the amplitude channel, producing high signal levels in both amplitude and phase.

Biological evaluation

Figure 4 shows reconstructions of *Saccharomyces cerevisiae* yeast recorded under VIS (λ = 561 nm) and DUV (λ = 300 nm) illumination using multi-distance acquisition mode. Each panel presents the full amplitude FOV reconstruction, with circular insets showing the separated amplitude (Amp) and phase (Ph) channels. The sample was mounted on a standard microscope slide under a coverslip. Because the sample preparation was not fully sealed, partial evaporation altered the sample between the VIS and DUV acquisitions. The DUV measurements were performed first, and the resulting evaporation effect is clearly visible in the left corners of the VIS FOV. Despite these differences, the contrast trends are clear: under DUV illumination, yeast exhibits strong amplitude contrast, while in VIS the signal is expressed predominantly in phase. This reflects increased light absorption and scattering, and higher phase sensitivity at shorter wavelengths, demonstrating how DUV illumination converts weakly scattering phase objects into strong mixed amplitude–phase signals.

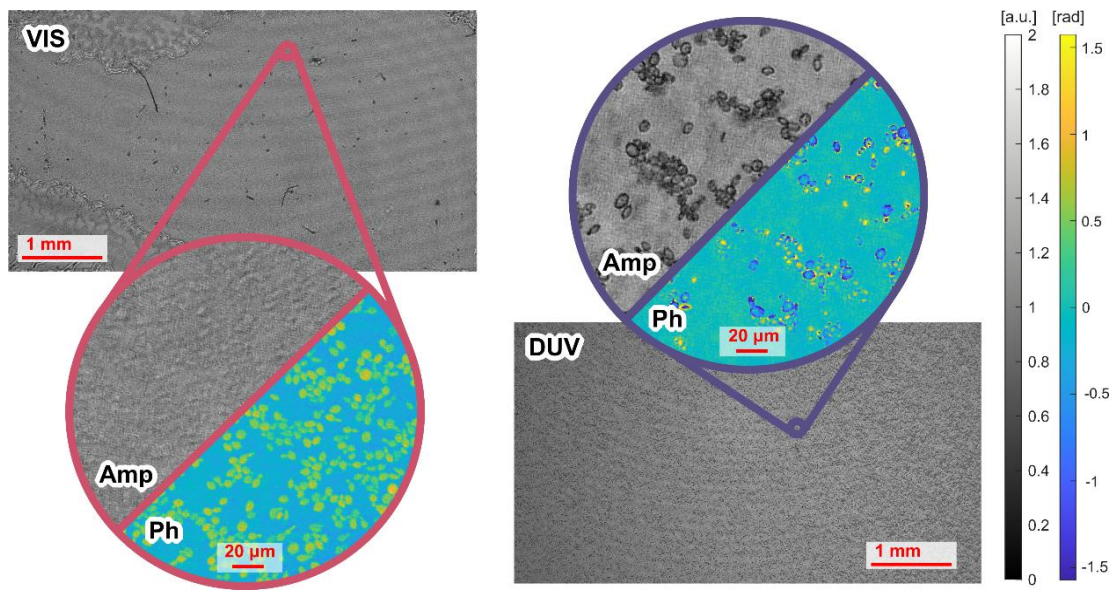

Figure 4. Comparison of PSR reconstructions of *Saccharomyces cerevisiae* yeast under VIS and DUV illumination. The full FOVs are shown in amplitude, while the circular insets display both amplitude and phase. Because yeast are phase objects, the VIS reconstructions exhibit none amplitude contrast, whereas in the DUV regime the cells appear with strong contrast in both amplitude and phase.

Figure 5 presents reconstructions of EVs obtained under VIS (λ = 500 nm) and DUV (λ = 260 nm) illumination. As in the yeast experiment, rectangular panels show the full amplitude FOV reconstruction, while the circular insets separate the amplitude (Amp) and phase (Ph) components. EVs were prepared in PBS between two coverslips and exhibited rapid drying between acquisitions. To enable a fair comparison (same refractive index difference), PBS was replenished before the VIS measurement, which introduced additional larger debris visible only in the VIS FOV. Despite these practical differences, the contrast behavior remains robust: under DUV illumination, EVs exhibit strong amplitude contrast mainly due to RNA absorption near 260 nm and increased light scattering, while in VIS they appear primarily as weak phase features.

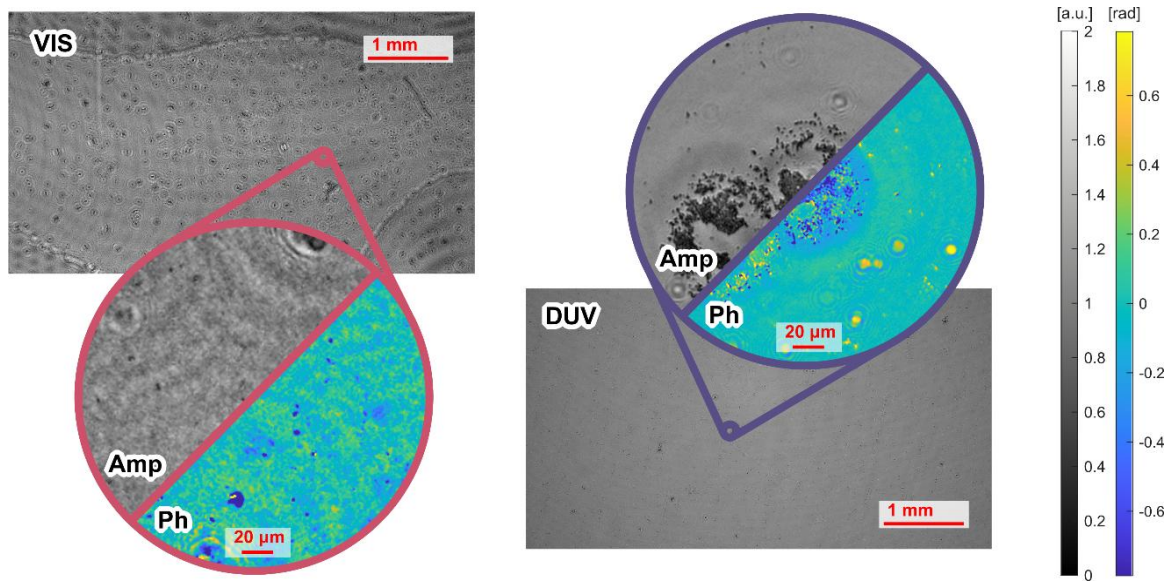

Figure 5. Comparison of PSR reconstructions of EVs under VIS and DUV illumination. The full FOVs are shown in amplitude, while the circular insets display both amplitude and phase. As EVs are weakly scattering phase objects in visible range, the VIS reconstructions provide negligible amplitude contrast and only faint phase signatures. In contrast, DUV illumination yields markedly enhanced visibility, with EVs in both amplitude (increased absorption of RNA and vesicle scattering) and phase channels.

We next imaged 30 µm thick unstained mouse liver sections using the lensless setup under both VIS and DUV illumination, Fig. 6. The large amplitude FOV of the tissue is shown at the top, with the marked region indicating the area enlarged below. Under VIS illumination ($\lambda = 650$ nm), the reconstructed amplitude image shows no clear subcellular features. In contrast, under DUV illumination ($\lambda = 330$ nm), numerous punctate structures become clearly visible in amplitude. Based on their morphology and distribution, we hypothesize that these features correspond to lipid-laden hepatic stellate cells (HSCs). HSCs (also known as Ito cells) are localized in the space of Disse between hepatocytes and sinusoidal endothelium. In their quiescent state, HSCs serve as reservoirs for vitamin A stored in lipid droplets. Morphologically, they display a stellate appearance with multiple cytoplasmic projections, and their nuclei have smooth contours[32]. Strong contrast in DUV comes from the selective absorption of retinoids, which have characteristic absorption bands near 330 nm. Such absorption is absent in the visible range, explaining why the same structures are not detectable in VIS amplitude. To validate our hypothesis, the same liver section was stained with Oil Red O (ORO) and re-imaged in the lensless system at a VIS wavelength of 632.8 nm. The resulting amplitude map reproduced the punctate distribution observed in the UV reconstruction. Independent Fourier ptychographic microscopy (FPM) measurement, performed using our custom-built system designed for high space-bandwidth product reconstruction[33,34], further confirmed the co-localization of stained lipids and unstained features observed in LH-DUV (see Fig. 7). Quiescent HSCs are readily visualized using ORO histochemical staining, where lipid (retinoid) droplets appear intensely red. ORO staining in the case of activated HSCs is very weak or not visible[35]. Together, these results indicate that LH-DUV imaging can reveal retinoid (lipid) accumulations without staining and reproduce the similar structures highlighted by ORO histochemistry features not detectable in VIS. This provides label-free high-throughput chemical specificity in histopathological application, here preliminarily applied - for the first time - to HSCs activation screening.

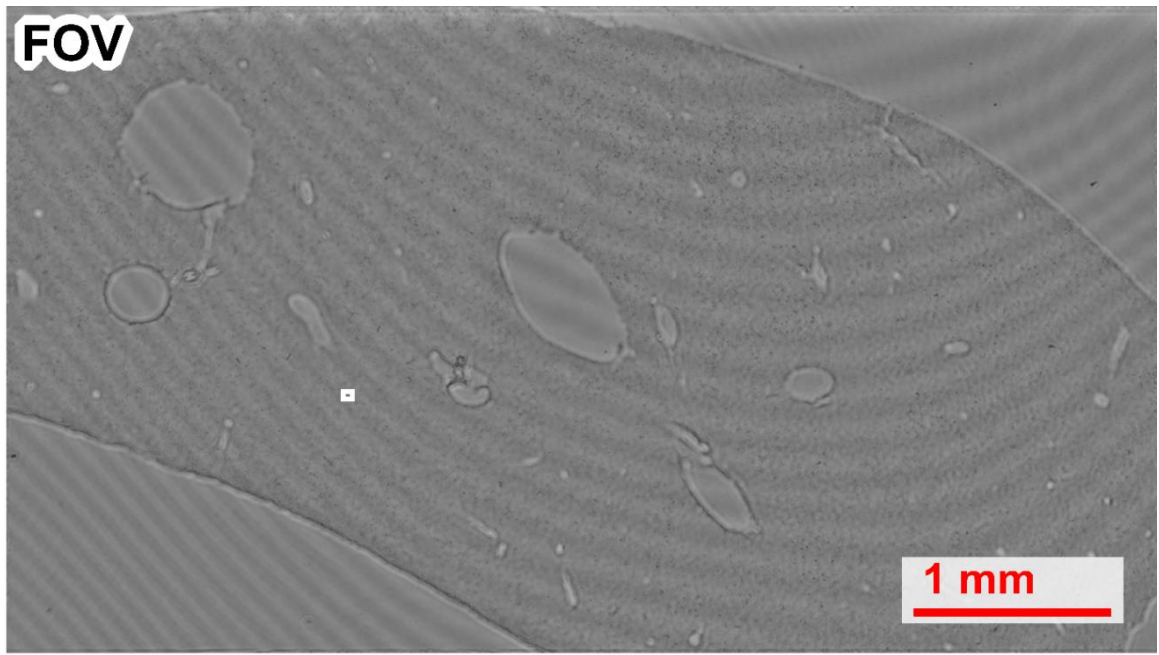
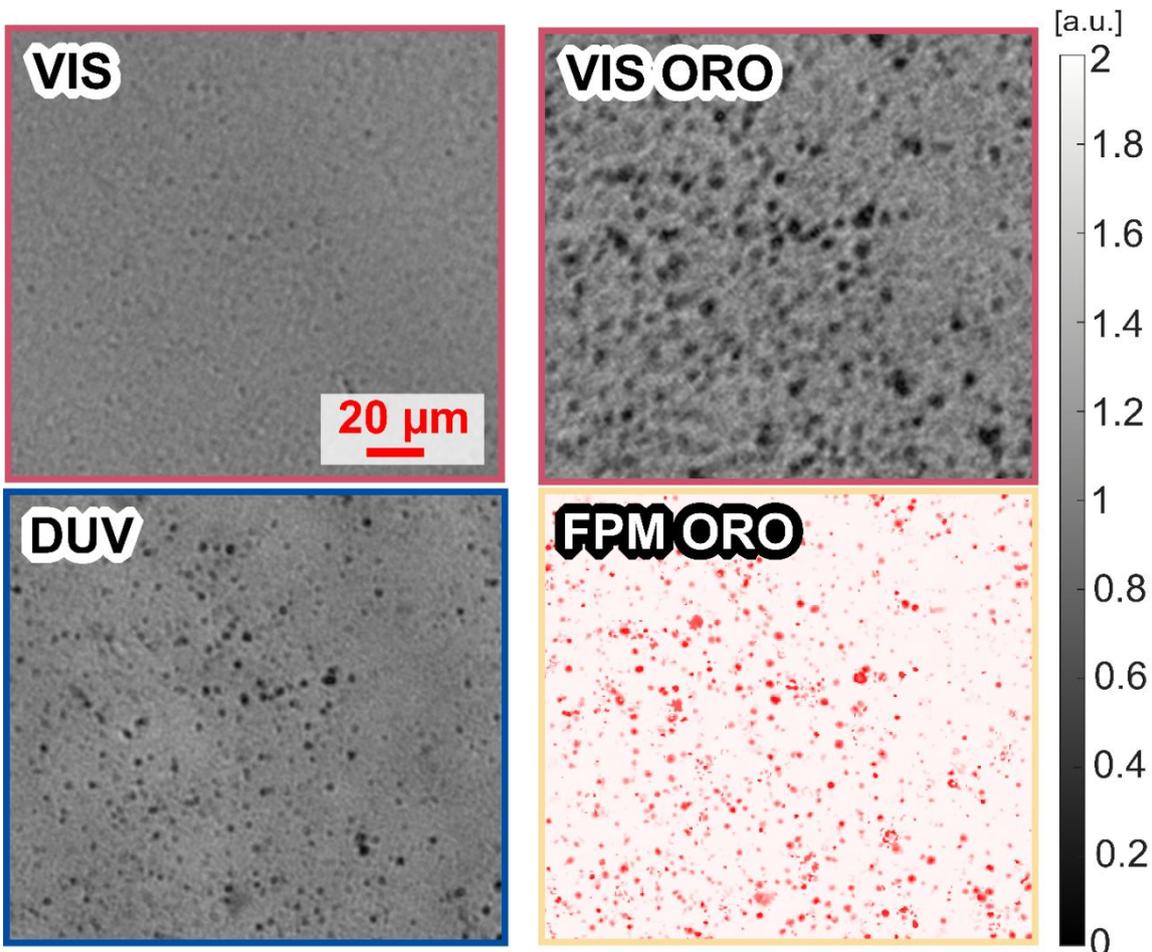

Figure 6. LH-DUV imaging of mouse liver tissue. The top panel shows the amplitude FOV reconstruction with the white square indicating the region magnified in the panels below. "VIS" and "DUV" correspond to lensless reconstructions obtained under VIS and DUV illumination, respectively, revealing punctate features in the DUV image that are absent in VIS. "VIS ORO" shows a lensless VIS reconstruction after ORO staining, highlighting lipid-rich droplets. Finally, "FPM ORO" presents a FPM reconstruction of the Oil Red O-positive stained sample, confirming the co-localization between retinoid-rich puncta detected in LH-DUV and lipid accumulations visualized by ORO staining, suggesting Ito cells distribution.

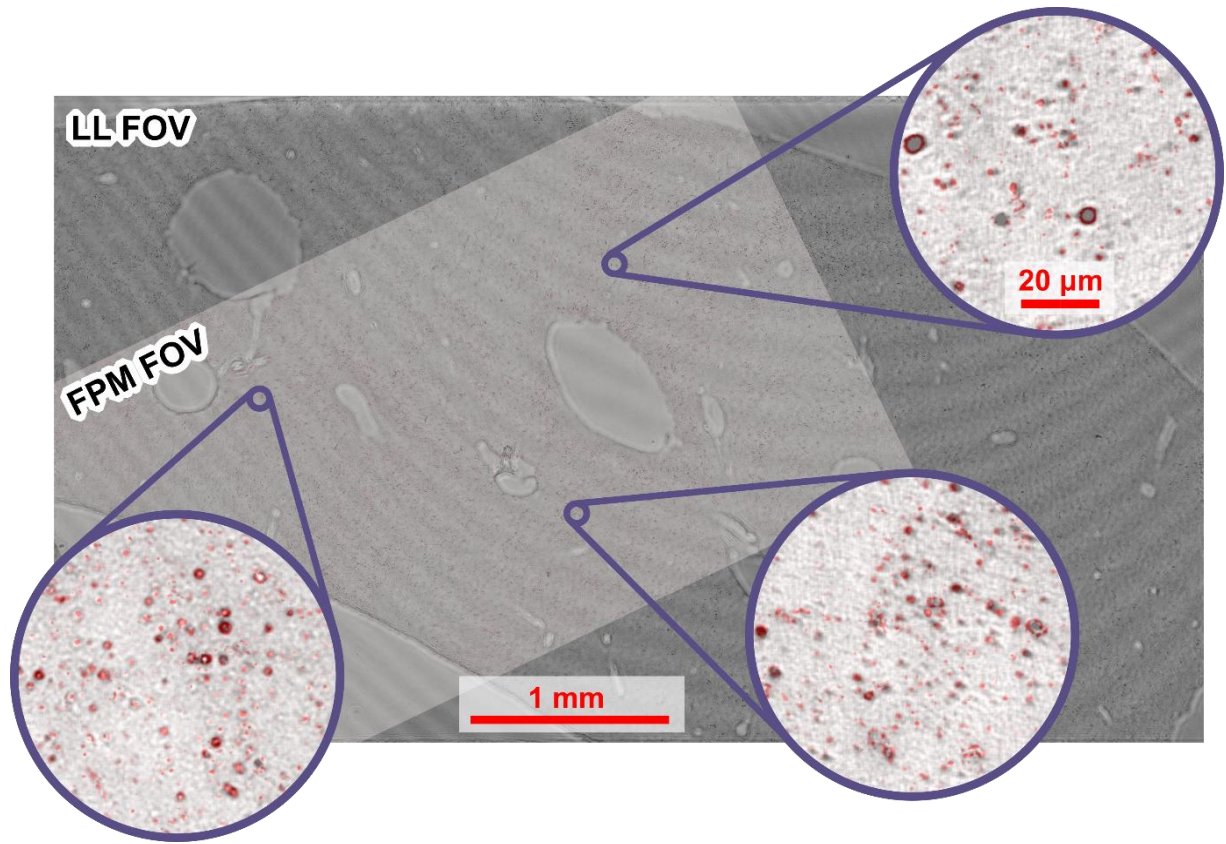

Figure 7. Co-localization of LH-DUV reconstructions with ORO-stained FPM images of mouse liver tissue. The large background panel shows the FOV obtained with the LH-DUV system (LL FOV), overlaid with the corresponding high-resolution FPM FOV. Circular insets highlight representative regions demonstrating a strong spatial correspondence between lipid-rich structures detected by the LH-DUV modality (rendered in grayscale) and the ORO-stained features visualized with FPM (marked in red). Across the entire overlapping FOV, lipid droplets exhibit consistent positional agreement in both modalities, confirming that DUV-induced contrast robustly identifies lipid-rich microstructures related to Ito cells without the need for chemical staining.

## Discussion

This work presents a first demonstration of chemical contrast in high-throughput label-free LH-DUV imaging. Many biomolecules including proteins, nucleic acids, and lipids, exhibit strong and selective absorption bands in the DUV. By tuning the illumination wavelength to these specific bands, it becomes possible to visualize and differentiate molecular components without extrinsic staining. This was exemplified by the detection of retinoid-rich hepatic stellate cells in unstained mouse liver tissue, where characteristic absorption at 330 nm produced strong amplitude contrast, revealing specific retinoid-rich structures that were effectively invisible under visible light illumination. For EVs - imaged for the first time here in DUV and without lenses - which are enriched with nuclear acids (small non-codding RNA mostly) their amplitude visibility was achieved at 260 nm within absorption characteristic for RNA molecules[36–38]. Taking into consideration that average EV dimension is in a range between 50 and 300 nm, one observes increased scattering in DUV which additionally amplifies holographic signal - a clear advantage of LH-DUV. The LL-DUV capabilities are particularly powerful because it eliminates the need for conventional staining procedures, which are often time-consuming, variable, and sample destructive. By enabling intrinsic chemical contrast directly from the specimen, LH-DUV provides a faster (large FOV and high-throughput), simpler, and non-invasive alternative to standard histological methods.

One of the main critical aspect of this work is the demonstration of quantitative lensless holographic DUV imaging under low dose (illumination). Gorti et al. previously studied that under DUV imaging conditions (255 nm, ~1 nW/μm², 50 ms exposure) more than 20,000 sequential acquisitions can be performed before reaching the photodamage threshold (1 μJ/μm²)[17]. In our experiments, the DUV illumination reaching the detector was below the minimum detection limit of our power meter (PM100D with S175C sensor, Thorlabs), whose active area is 18 mm ×18 mm. Even if the minimum detectable power (100 μW) were used in this calculation, it would correspond to a surface power density of approximately 0.00031 nW/μm² - already three orders of magnitude lower than threshold that was considered as safe for imaging in the mentioned work[17]. Since the signal remained below the detection floor, the actual delivered dose was actually much lower. This provides strong evidence that our illumination levels are orders of magnitude below reported cytotoxic DUV doses, supporting the compatibility of the proposed LH-DUV imaging technique with sensitive or live specimens.

A central numerical contribution of this work is the implementation of PSR strategy specifically adapted to low illumination intensity in LH-DUV by supporting it with TV denoising and robust darkfield driven autofocusing[40]. The algorithm operates seamlessly in both multi-distance and multi-wavelength acquisition modes, offering complementary advantages. The multi-distance mode relies on a single fixed-wavelength laser and simple axial translations, making it particularly well suited for standard laboratory environments where tuneable sources may not be available. By contrast, the multi-wavelength mode requires no mechanical motion and is ideal for studies where the imaging system must remain completely static, for example, when imaging live delicate biological samples. The robustness of the reconstruction process under both modes stems from the use of a global Fourier transform framework (central to the hologram propagation - numerical refocusing - procedure), which effectively exploits information redundancy across the full sensor area. In other words, each point in the object plane scatters onto the entire sensor creating a hologram, while global Fourier transform uses all pixels in the hologram (image) domain to calculate each pixel in the spectral domain and allow for propagation (refocusing). These features enable the holographic reconstruction algorithm to remain stable even under extremely low signal conditions, as the large number of pixels is more important than detected intensity dynamic range in each pixel.

Despite these advantages, several limitations warrant consideration. First, the dose was indirectly controlled via exposure time because the incident power was below meter sensitivity; future work should calibrate absolute radiant exposure and photons-per-pixel to enable quantitative dose-response analysis. Second, usable wavelengths depended on substrate transmission (standard microscope glass absorbs strongly at ~300 nm and entirely blocks shorter DUV wavelengths); systematic evaluation of common substrates and the adoption of fused silica or quartz coverslips could help extend the accessible spectral range, albeit at increased system cost. It should also be noted that not only the substrate material but also the immersion medium can strongly absorb DUV light, so it should be chosen carefully. While the achieved lateral resolution of 870 nm in proposed LH-DUV method already provides high-quality submicron imaging, it does not yet reach the theoretical diffraction limit set by the illumination wavelength. Because DUV illumination inherently supports higher resolution, using a detector with a smaller pixel pitch would make it possible to overcome the current sampling bottleneck and approach the wavelength-limited resolution of the system. At present, the main constraint lies in the fabrication of sensors with insufficiently small pixels, which can be further numerically reduced in size by 2 via PSR.

All presented results position novel LH-DUV imaging as a powerful and practical modality for quantitative, chemically specific, label-free bioimaging at extremely low illumination levels in large FOV deploying regular VIS CMOS sensors. Its combination of cost efficiency, optical simplicity, and robust reconstruction makes it a promising platform for large-area tissue imaging, high-throughput bio-chemical screening, and future live-cell studies, with potentially impactful applications in biomedicine, biophotonics and disease diagnostics.

## Methods

### Setup details

The system employed plane-wave illumination to ensure uniform magnification[39] across all recorded holograms. The tunable laser used in most experiments was the Ekspla NT242 (programmable range 210-2100 nm, tuning resolution 1 nm, FWHM linewidth 3-5 cm$^{-1}$, pulse duration 3-6 ns, repetition rate 1 kHz). The wavelength selection was achieved by fine adjustments inside the optical parametric oscillator resonator, allowing discrete tuning steps. To validate the applicability of the approach using simpler light sources, a fixed-wavelength DUV laser (Teem Photonics SNU-02P-100, $\lambda = 266$ nm, pulse duration ≈ 550 ps, repetition rate ≈ 6 kHz, output energy > 0.3 μJ) was employed in the low-illumination CMOS validation experiment described in Results section (Fig. 1).

In the experiments, we employed two different VIS board-level CMOS cameras: Allied Vision Alvium 1800 U-2050m Bareboard (5496 × 3672 pixels, 2.4 μm pixel size) and iDS uEye+ U3-38J2XLE Rev. 1.2 (3864 × 2176 pixels, 1.45 μm pixel size), to demonstrate the versatility of the method. Although both cameras are specified for the visible spectral range, they exhibited sufficient sensitivity in the DUV, allowing the use of standard, low-cost detectors without requiring specialized DUV-optimized sensors. This is possible because board-level CMOS sensors are used without front-end optics, so the incident DUV light directly reaches the photodiode layer. Apart from a thin protective cover plate, typically made of silicon or fused silica with good DUV transmittance, no additional optical elements are present to attenuate the signal. This simple configuration ensures that the quantum efficiency, although reduced compared to VIS, remains sufficient for lensless holographic reconstruction. According to the available data sheets, the quantum efficiency of the Allied Vision camera is approximately 5.7% at 320 nm, while that of the iDS camera reaches about 37% at 400 nm.

Importantly, unlike conventional lens-based systems, the LH-DUV image reconstruction process relies on a global transformation of the entire raw image captured by the camera, enabling the recovery of object information even under low DUV intensity conditions. Specifically, the key element of LH-DUV reconstruction is the Fourier transform, which is a global operation-meaning that all pixels in the image domain contribute to the calculation of each pixel value in the Fourier domain, and vice versa. Likewise, in LH-DUV, each object point is redundantly encoded by many sensor pixels, allowing robust recovery of object information despite low intensity levels in individual pixels. This combination of direct sensor illumination, absence of front-end optics, and global phase retrieval makes the configuration both cost-efficient and technically powerful, crucially for exotic and powerful DUV regime.

### Pixel-super resolution lensless holographic reconstruction

We employed a multi-frame strategy implemented in two alternative acquisition modes: (i) axial multi-distance, in which a set of holograms are recorded at various propagation distances using a single illumination wavelength, and (ii) multi-wavelength, in which holograms are successively recorded in a motion-free fashion at different illumination wavelengths. In both modes, the resulting sequence of holograms - each with distinct fringe patterns with spatially shifted Gabor fringes - provides the sub-pixel diversity required to surpass the native sensor pitch limitation. For reconstruction, the two modes are operationally equivalent: the same registration and PSR pipeline is applied, differing only in the source of diversity (z-distance versus wavelength). The use of plane-wave illumination and the short sample-to-sensor distance ensures identical magnification across all frames, regardless of the PSR mode used.

In the multi-distance mode, the *z*-stack of 20 holograms was acquired with a motorized linear stage (Standa 034223), with the first hologram recorded at the closest attainable position (around 1-2 mm) and subsequent holograms obtained in 50 μm axial steps. In the multi-wavelength mode, the geometry was kept fixed, and diversity was introduced by tuning the laser wavelength, resulting in 22 holograms acquired at distinct wavelengths (3 nm wavelength step). In this simple unity-magnification geometry, the accessible FOV is up to approximately 116 mm², which is equal to camera sensor dimensions (for Allied Vision camera).

The PSR reconstruction relies on an iterative multi-frame phase retrieval procedure, which exploits the sub-pixel information encoded in a series of holograms acquired with either multi-distance or multi-wavelength diversity. As part of our implementation, we introduce a TV denoising step, which substantially reduces noise in heavily noise-contaminated measurements - a regime characteristic of LH-DUV acquisition, where extremely low photon budgets can still yield recoverable information. Moreover, we implement robust computational darkfield-driven (thus both amplitude and phase object sensitive) autofocusing[40] which is crucial for multi-hologram alignment. These original additions improve reconstruction under extremely low-SNR conditions. The entire new processing pipeline, schematically illustrated with Fig. 2, is explained with the pseudocode below.

Algorithm 1

|   | PSR algorithm |
|---|---|
|   | Inputs: $I(x, y, n); z(n); \lambda; \Delta; \tau; \alpha$ |
|   | Outputs: $E(\hat{x}, \hat{y}, z = 0)$ |
| 1 | $E(\hat{x}, \hat{y}, z(1)) = \left[\sqrt{I(x, y, 1)}\right]_S^{\hat{S}}$   % Initial guess -> upsampled amplitude |
| 2 | for: $j = 1:J$ %Continue for J global iterations |
| 3 |   for: $n = 1:N$ % Interate over holograms |
|   |   % Propagate from $n$-th hologram to sample plane |
| 4 |   $E(\hat{x}, \hat{y}, 0) = propagate(E(\hat{x}, \hat{y}, z(n)), -z(n), \lambda, \Delta/\tau)$ |
| 5 |   $E'(\hat{x}, \hat{y}, 0) = TV(E(\hat{x}, \hat{y}, 0))$  % TV denoising |
| 6 |   $n_{+1} = n + 1$   % Index of subsequent image |
| 7 |   if: $n_{+1} == N$ |
| 8 |     $n_{+1} = 1$ |
|   |   % Propagate from sample to $n_{+1}$-th hologram plane |
| 9 |   $E(\hat{x}, \hat{y}, z(n_{+1})) = propagate(E(\hat{x}, \hat{y}, 0), z(n_{+1}), \lambda, \Delta/\tau)$ |
| 10 |   $A(x, y) = \lfloor\|E(\hat{x}, \hat{y}, z(n_{+1}))\|\rfloor_{\hat{S}}^{S}$   % Downsample amplitude |
| 11 |   $B(\hat{x}, \hat{y}) = \left[\frac{\sqrt{I(x,y,n_{+1})}}{A(x,y)}\right]_S^{\hat{S}}$   % Impose intensity constraint and upsample |
|   |   % Actualize optical field in $n_{+1}$ plane |
| 13 |   $E(\hat{x}, \hat{y}, z(n_{+1})) = (1 - \alpha) \cdot E(\hat{x}, \hat{y}, z(n_{+1})) + \alpha \cdot B(\hat{x}, \hat{y}) \cdot E(\hat{x}, \hat{y}, z(n_{+1}))$ |
|   | % Final backpropagation |
| 14 | $E(\hat{x}, \hat{y}, 0) = propagate(E(\hat{x}, \hat{y}, z(1)), -z(1), \lambda, \Delta/\tau)$ |

Throughout the pseudocode, $(x, y)$ stand for the registered images (original size) coordinates, while $(\hat{x}, \hat{y})$ denote the coordinates in the reconstructed image (upsampled) domain. The $[a]_S^{\hat{S}}$ denote the upsampling operation of matrix $a$ from $S$ (original dimensions) to $\hat{S}$ (reconstruction dimensions) image size, while $\lfloor a \rfloor_{\hat{S}}^{S}$ stands for downsampling of $a$ from $\hat{S}$ to $S$ image size. The $I(x, y, n)$ represent the $n$-th hologram ($n = 1, ..., N$), recorded in an in-line lensless geometry, registered with $n$-th camera-sample distance ($z(n)$; constant for multi-lambda reconstruction) or $n$-th illumination wavelength ($\lambda(n)$; constant for multi-distance reconstruction). For each frame, the object-camera distance $z(n)$ was estimated automatically by maximizing a dark-focus-based sharpness metric[40] (illustrated in Fig. 8 within Preprocessing). In multi-distance mode, the known mechanical step of stepper motor allowed us to bound the axial search range reducing runtime. The PSR takes as input the wavelength $\lambda$, an upsampling factor $\tau$ (fixed to 4 in all presented results, since larger values (e.g. $\tau$=8) did not provide further resolution enhancement), and a relaxation factor α that controls the update strength (here α = 0.2).

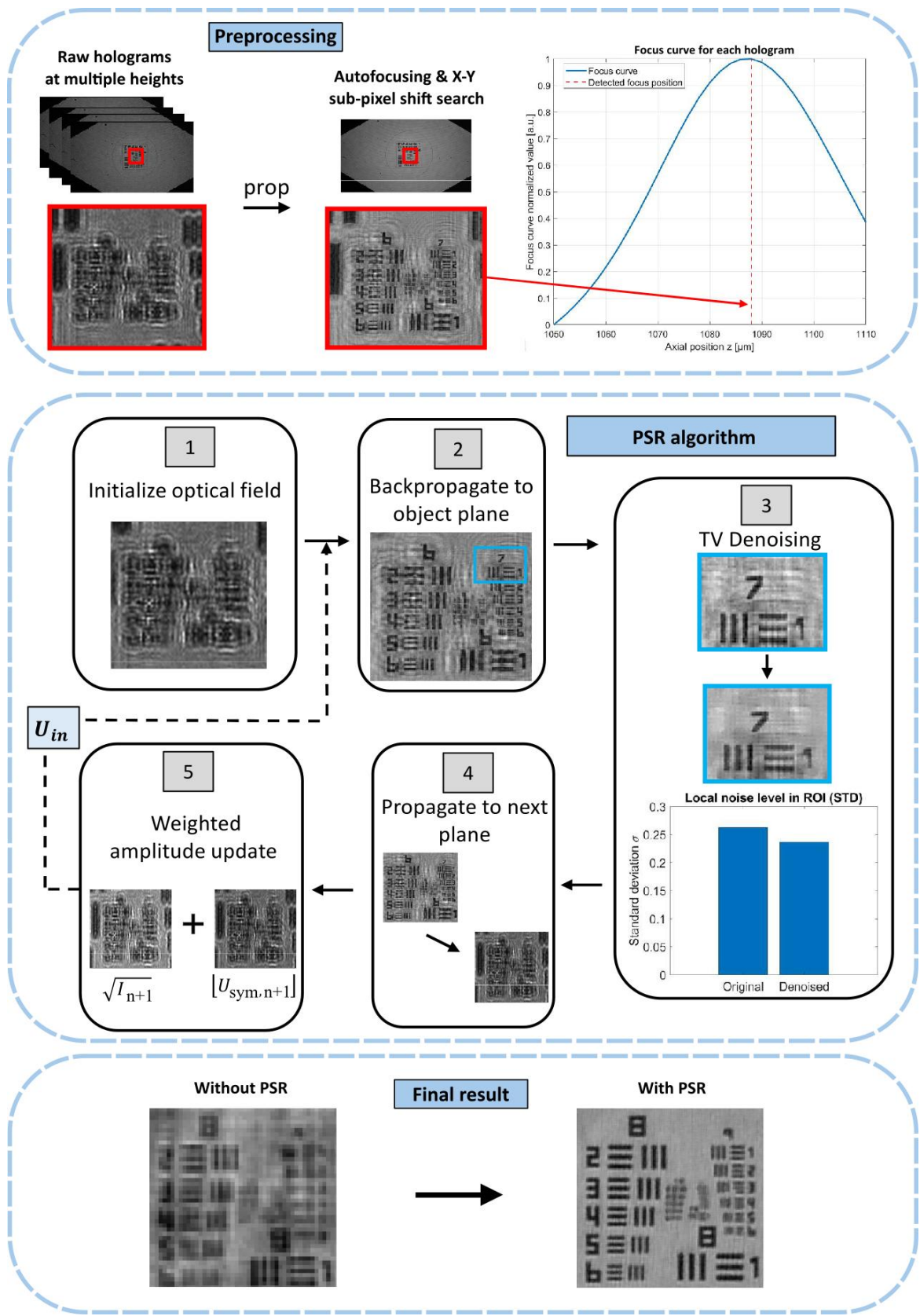

Figure 8. PSR pipeline: preprocessing (autofocusing), algorithm (iterative reconstruction steps (1–5)) and final enhanced resolution reconstruction.

The PSR routine begins by upsampling the first hologram from the native camera dimensions $S$ to the final reconstruction dimensions $\hat{S}$, seeding a high-resolution estimate (Step 1 in Fig. 8). Then J iterations of the algorithms begin. In each iteration the field is back-propagated to the object (sample) plane (Step 2 in Fig. 8), after which TV denoising suppresses noise and twin-image artifacts while maintaining sharp object features (Step 3 in Fig. 8). Subsequently, the complex wavefront is propagated forward to the next sensor plane (Step 4 in Fig. 8); its amplitude is downsampled to the camera grid, replaced by the measured amplitude $\sqrt{I_{k+1}}$, and upsampled back to $\hat{S}$ while preserving phase. The estimate is updated using a relaxation weight $\alpha$ (Step 5 in Fig. 8), and the cycle continues across all N frames. After $J$ passes, a final propagation to the object plane yields the reconstructed complex field.

In practice, the optimal strength of the TV denoising step should be adjusted according to the noise level.

Sample preparation – yeast

*Saccharomyces cerevisiae* WUT3 yeast from the Warsaw University of Technology (https://www.wutyeastcollection.pw.edu.pl/) was seeded onto Sabouraud (SAB) agar medium (Biomaxima, Lublin, Poland) and incubated for 24 h at 37°C. Then, single colonies were used to inoculate 10 mL of SAB broth (Biomaxima, Poland) and incubated overnight at 37°C with shaking at 240 rpm in a Benchtop Shaker SI-600R Lab Companion (Jeio Tech, South Korea). 1 mL of the overnight culture was taken and centrifuged at 3,000×$g$ for 3 min. The supernatant was discarded. The cell pellet was then resuspended in 1 mL of PBS (phosphate-buffered saline without calcium or magnesium, VWR International, Belgium). The sample was spun down again at 3,000×$g$ for 3 min. This step was repeated to wash away any remaining medium from the pellet. Finally, the cells were suspended in 1 mL of PBS. Then, 750 μL of the suspension was transferred to a new tube and 250 μL of 10% paraformaldehyde (w/v, Sigma-Aldrich, Germany) was added for fixation. The samples were gently mixed for 30 min, then centrifuged and rinsed three times with PBS as described above. Finally, the cell pellet was resuspended in 1 mL of PBS, and 5 μL of the cell suspension was applied to a microscope slide (VWR® Polysine®, Adhesion Slides Catalog # 631-0107V) and sealed with a cover glass (VWR®, Cover Glasses, Square Catalog #631-1568).

Sample preparation – EVs

EVs were obtained from adipose mesenchymal stem cells (AMSC) (Cat No: PB-CH-642-0511-1, PELOBiotech GmbH, Germany) between 3rd and 4th passage. AMCS were cultured with xeno-free stem cell medium kit (Cat No:PB-C-MH-675-0511-XF, PELOBiotech GmbH, Germany) enhanced with 25 ml growth supplement and after 3rd passage medium was changed to the mixture 1:1 of high glucose DMEM with GlutaMAX™ (Cat No: 61965026, Gibco) and 10% of fetal bovine serum (FBS) (Cat No: FBS-H1-11A, Capricorn, GmbH, Germany). AMSC derived EVs were isolated after 24 hours incubation with serum free medium and characterized according to the differential centrifugation protocol at final 18,000 g sedimentation and filtration through a 1000 kD MWCO membrane (Cat No: 131486) procedures[41,42].

Sample preparation – mouse liver

Mice: Liver was obtained from inbred C57BL/6J mice (bred at the Center for Experimental Medicine in Białystok and the M. Mossakowski Medical Research Center in Warsaw). The animals were provided a controlled environment (temperature 24°C, 12/12 light/dark cycle) with ad libitum access to water and feed. The animals were euthanized by cervical dislocation following exposure to isoflurane (FDG9623, Baxter). Isolated tissues were fixed for 24 h in a 4% paraformaldehyde solution (BD Cytofix/Cytoperm, No. 554722, BD Biosciences) at 4°C before being used in experiments. All procedures were conducted in accordance with the Directive of the European Parliament and Council No. 2010/63/EU on the protection of animals used for scientific purposes.

Preparation of tissue sections: The fixed liver was embedded in a 3% (w/v) agarose solution (Sigma–Aldrich) in water and subsequently sectioned to the 30 μm thickness using a Vibratome (VT1000S,

Leica). The sections were stored in 1× PBS (Sigma–Aldrich) supplemented with 0.05% sodium azide (NaN3, Sigma–Aldrich).

Tissue optical clearing and staining: Ce3D was prepared according to a published protocol[43]. In brief, a 40% (v/v) solution of N-methylacetamide (M26305-100G, Sigma–Aldrich) in PBS was prepared and used to dissolve Histodenz (D2158-100G, Sigma–Aldrich) to 86% (w/v) at 37°C. After complete dissolution, achieved in 5–6 h, Triton X-100 (0.1%, v/v, Sigma–Aldrich) and 1-thioglycerol (0.5%, v/v, M1753-100ML, Sigma–Aldrich) were added to the solution. For clearing the 30 μm liver sections, no fixed incubation period was used; instead, the samples remained immersed in the Ce3D. Oil Red O 0.4% stock solution was prepared by dissolving Oil Red O (O-0625, Sigma–Aldrich) in 100% isopropanol. The working solution was obtained by mixing three parts of the stock solution with two parts of double-distilled water and filtering through a 0.2 μm membrane after 20 minutes.

For co-locailization, liver sections where first cleared and imaged with Ce3D, washed with PBS, immersed in working solution of Oil Red O for 10-15 minutes, washed 4 times with PBS, immersed briefly in Ce3D and re-imaged.

## Funding

Funded by the European Union (ERC, NaNoLens, Project 101117392). Views and opinions expressed are however those of the author(s) only and do not necessarily reflect those of the European Union or the European Research Council Executive Agency (ERCEA). Neither the European Union nor the granting authority can be held responsible for them.

The research was carried out using equipment co-funded by the Warsaw University of Technology within the Excellence Initiative: Research University (IDUB) program.

Experiments with EVs were supported with the grant to Prof. E. Stępień (2022/45/B/NZ7/01430) by the National Science Center (NCN). Experiments with yeast were supported with the grant to Prof. Jolanta Mierzejewska (2023/49/B/NZ9/03663) by the National Science Center (NCN).

## Data availability

Data underlying the results presented in this paper are not publicly available but may be obtained upon reasonable request.

## Authors contribution

M.T. conceived and supervised this project and provided funding and resources for the project. M.T., P.A., M.R. designed the research. P.A, P.Z., E.W. and G.S. performed the measurements and acquired the data. M.R., K.N. and J.W. developed the PSR+TV algorithm. P.A., K.N., M.R. and J.D. performed the numerical reconstructions. P.A. generated the manuscript figures. P.M. and A.Ch. provided and prepared mouse liver sample. J.M. provided the yeast sample. E.S. provided the EVs sample. M.L. performed histopathological analysis with additional staining and imaging by P.A.. M.T., P.A. and M.R. wrote the manuscript and all authors commented on the manuscript.

**Competing interests**

The authors declare no competing interests.

**Code availability**

The reconstruction algorithm used in this study is provided in the Methods section.

**Supplementary Information**

Supplementary Information is available for this paper in Supplementary Document 1.

# Supplementary Document 1

**Low photon budget DUV lensless holographic (LH-DUV) reconstruction**

In this supplementary experiment, we quantitatively evaluated the influence of single-hologram angular spectrum (AS) propagation, multi-frame pixel super-resolution (PSR) reconstruction, and PSR augmented with total-variation (TV) denoising on the noise level in LH-DUV reconstructions. Noise was quantified as the standard deviation (STD) of the intensity within a background region of a USAF resolution target (an area without any object features). For PSR and PSR with TV, the STD was computed from 1,674,240 pixels, whereas for AS the number of pixels was four times smaller due to the upsampling factor inherent to the PSR algorithm. All results summarized in Table 1 correspond to the same datasets used for Figure 2 of the main manuscript.

As shown in Table 1, the exposure time of 0.3 ms - representing the most challenging case, in which a single hologram contains only three effective gray levels - yields the highest noise level when reconstructed with AS. Applying PSR to these data reduces noise by approximately 30%, which already constitutes a substantial improvement arising from multi-frame averaging. When TV denoising is additionally incorporated, the noise level decreases by a further factor of three relative to PSR alone. For an exposure of 0.5 ms, a similar trend is observed: PSR reduces noise by about 30% (from 25.41 to 17.40), and the application of PSR with TV yields an additional ~40% reduction. The same relationship persists for the 1 ms dataset (third row), where PSR again provides an initial ~30% noise reduction, followed by a further three-fold decrease when TV denoising is applied. Importantly, this dataset also achieves the final spatial resolution obtained under standard recording conditions (long exposure times).

The final row of Table 1 presents results for long-exposure, high-quality data containing approximately 150 gray levels. Under these conditions, PSR reduces noise by ~40%. As expected, the contribution of TV denoising is smaller here (approximately 14%), which reflects the already low intrinsic noise level of the long-exposure high-photon-budget raw data. Nonetheless, it should be emphasized that the proposed algorithm was specifically designed to suppress noise in extremely challenging low-photon-budget conditions - scenarios in which phototoxicity is a concern and illumination reaching the detector must be minimized to avoid biological damage. Despite this specialization, the method remains effective when applied to standard, high-quality datasets.

Overall, these results demonstrate that while PSR alone consistently decreases noise across all illumination conditions, the incorporation of TV denoising provides a additional benefit for extremely noisy holograms, reducing noise by up to a factor of three while preserving spatial resolution, as also illustrated in Figure 2 of the Results section. Proper selection of TV parameters is essential to match the noise level of the data; when chosen appropriately, PSR with TV achieves substantial noise suppression without compromising image detail. As the strength of TV denoising is adjusted to the data at hand calculated STD values are heterogenous with respect to the exposure time. In other words, column PSR with TV in Table 1 has no intrinsic significant trend due to manual adjustment of the TV parameters. However, all values are significantly lower in comparison with both remaining columns which quantitatively assesses the proposed method as favourable.

Table 1. Comparison of noise levels in reconstructed amplitude images for four representative exposure times and three reconstruction strategies. The AS method, based on single-frame back-propagation, exhibits the highest noise across all exposure settings, reflecting its sensitivity to low photon counts. Multi-frame PSR substantially reduces noise, while PSR combined with TV denoising provides the most robust performance.

|  | AS | PSR | PSR with TV |
|---|---|---|---|
| Exp. Time 0.3 ms | 31.50 | 24.03 | 7.91 |
| Exp. Time 0.5 ms | 25.41 | 17.40 | 10.16 |
| Exp. Time 1 ms | 18.04 | 12.39 | 4.77 |
| Exp. Time 15 ms | 11.12 | 6.94 | 6.13 |

To further investigate the impact of signal level on reconstruction noise, we conducted an additional experiment in which the exposure time was progressively reduced to systematically decrease the signal-to-noise ratio. In these measurements, we used a tunable laser and an Allied Vision camera. The exposure time was stepwise decreased starting from the longest acquisition (1800 ms) - initially in 200 ms intervals down to 200 ms, and subsequently in finer 50 ms increments down to 100 ms. This approach enabled us to evaluate how noise evolves as the detected signal becomes progressively weaker.

Figure S1 shows, with blue points, how the number of distinct gray levels in the hologram histograms decreases with shorter exposure times. Noise levels were quantified via STD within a background region of the USAF target devoid of structural features, and the corresponding standard deviations are plotted for both reconstruction methods (AS and PSR with TV). Single-frame AS results are marked with "×", whereas PSR with TV results are marked with "*". For exposure times below 100 ms, the recorded signal collapsed into only two gray levels, becoming dominated by noise and unsuitable for meaningful reconstruction.

A noteworthy observation is that the noise level remained relatively stable down to an exposure time of approximately 600 ms, which corresponds to histograms containing several gray levels (~10). Within this range, both reconstruction approaches exhibited stable, method-specific noise levels - AS retaining its characteristic noise and PSR with TV maintaining substantially lower noise - despite the strongly reduced optical power reaching the detector. This underscores the potential of LH-DUV to image biological samples under low-illumination conditions without introducing significant photodamage, while still maintaining reconstruction quality comparable to data acquired under standard illumination (both in terms of resolution and signal-to-noise ratio).

When the signal was further reduced by shortening the exposure time below this threshold, the noise increased exponentially. Nevertheless, PSR with TV continued to provide substantial noise suppression even under extremely low-photon-budget conditions. It should be emphasized, however, that in the most extreme cases - such as holograms containing only three gray levels and similar borderline conditions - the achievable reconstruction resolution becomes limited by the noise floor itself.

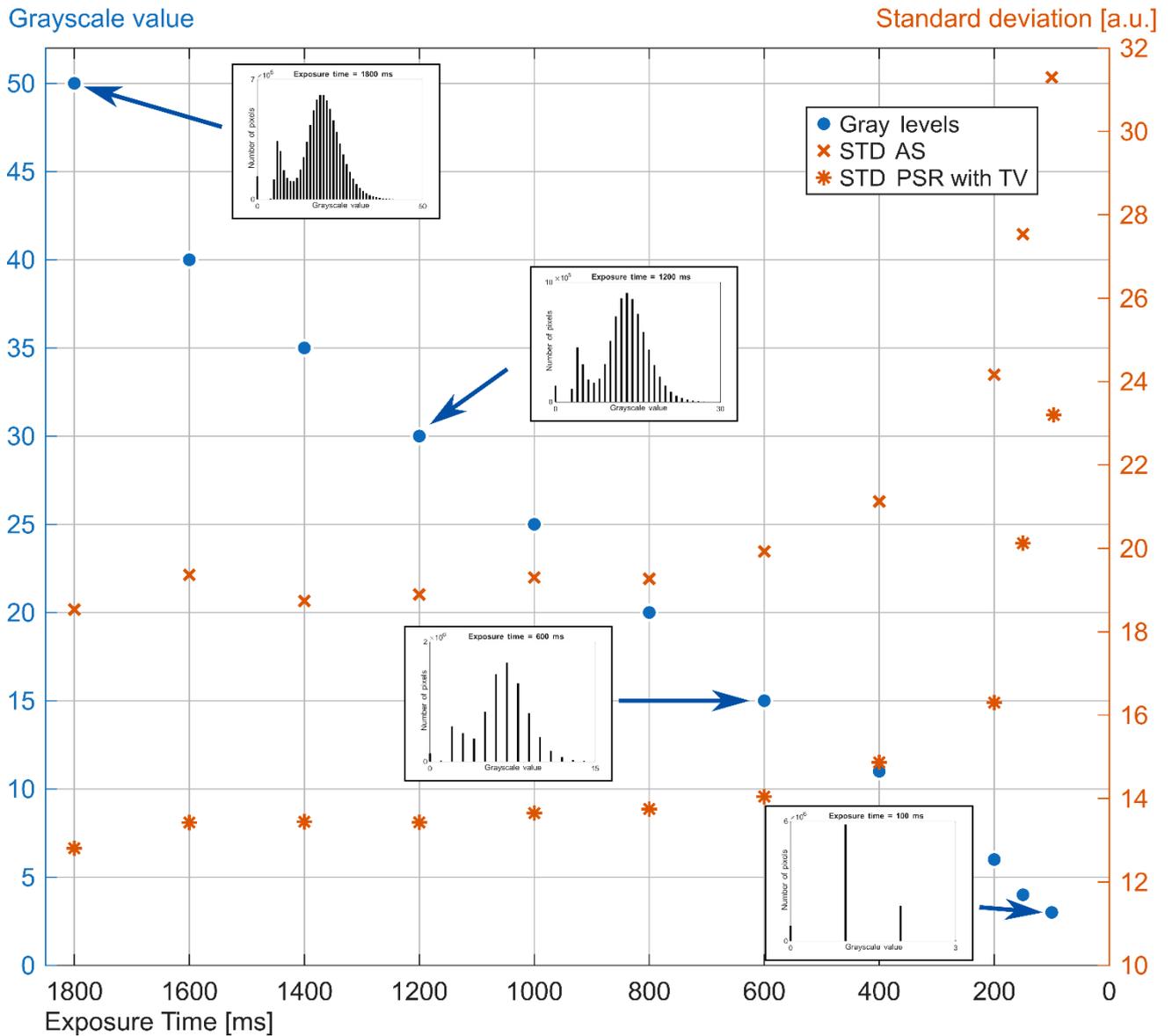

Figure S1. Relationship between exposure time, effective grayscale depth of a single hologram, and noise levels in LH-DUV reconstructions.

These experiments collectively demonstrate that LH-DUV holography is remarkably robust across detectors and operates effectively even when using standard CMOS cameras originally designed for the visible range. Although the two CMOS cameras used in this study exhibited markedly different maximum exposure times and dynamic ranges and quantum efficiencies in the DUV, meaningful reconstructions were achievable in both cases, despite the fact that that all measurements were performed deep within the low illumination regime (much below 100 uW measurable by power meter PM100D with S175C sensor, Thorlabs). This further highlights the inherent efficiency of LH-DUV and its natural resistance to phototoxicity. Nevertheless, the ability to obtain high-quality amplitude and phase reconstructions using detectors originally designed for the visible range underscores the unique practicality and versatility of the LH-DUV approach.